\documentclass{article}
\usepackage{spconf,amsmath,graphicx}
\usepackage[dvipsnames]{xcolor}
\usepackage{subcaption}
\usepackage{url}
\usepackage{hyperref}
\usepackage{fancyhdr}

\title{Towards Streaming Speech-to-Avatar Synthesis}
\name{Tejas S. Prabhune, Peter Wu, Bohan Yu, Gopala K. Anumanchipalli}
\address{University of California, Berkeley}
\begin{document}
\fancypagestyle{copyright}{\fancyhf{}\renewcommand{\headrulewidth}{0pt}\fancyfoot[L]{© 2024 IEEE. Personal use of this material is permitted. Permission from IEEE must be obtained for all other uses, in any
current or future media, including reprinting/republishing this material for advertising or promotional purposes, creating new
collective works, for resale or redistribution to servers or lists, or reuse of any copyrighted component of this work in other
works.}}

\thispagestyle{copyright}

\maketitle
\begin{abstract}

Streaming speech-to-avatar synthesis creates real-time animations for a virtual 
character from audio data. Accurate avatar representations of speech are 
important for the visualization of sound in linguistics, phonetics, and 
phonology, visual feedback to assist second language acquisition, and 
virtual embodiment for paralyzed patients. Previous works have highlighted the 
capability of deep articulatory inversion to perform high-quality avatar 
animation using electromagnetic articulography (EMA) features. However, 
these models focus on offline avatar synthesis with recordings rather than 
real-time audio, which is necessary for live avatar visualization or 
embodiment. To address this issue, we propose a method using 
articulatory inversion for streaming high quality facial and inner-mouth 
avatar animation from real-time audio. Our approach achieves 
130ms average streaming latency for every 0.1 seconds of audio
with a 0.792 correlation with ground truth articulations. 
Finally, we show generated mouth and tongue animations to demonstrate the 
efficacy of our methodology.

\end{abstract}
\section{Introduction}
\label{sec:intro}

Speech-driven avatar animation is useful for many applications in speech and linguistics. 
Specifically, it can facilitate second language (L2) pronunciation learning via visual feedback 
\cite{sla2015, emavisual2010, ultra2008} and aid hearing-impaired individuals to lip-read when only an 
audio signal is available during communication. In addition, accurate facial and 
tongue animation has been shown to help with virtual embodiment for paralyzed patients
\cite{bravo2023}. Solutions to the speech-driven avatar task date back to 
\cite{synface2003, synface2009}, which proposed predicting phonemes using a 
combination of Qualisys optical motion tracking and electromagnetic articulography (EMA) data.

Key tasks within the development of automated avatars include both real-time and offline speech-driven 
animation of the face and inner mouth, as needed in interactive systems or multimedia like video 
games \cite{nutshell2006}. Previous works focusing on offline synthesis have achieved success using 
face scans \cite{voca2019, voice2face} as well as using various input modalities 
like MRI \cite{mri2000} and/or EMA \cite{mriema2017} to model the movement of articulators. More 
recently, deep articulatory inversion techniques have shown to produce high-quality speech-to-EMA 
models and subsequent avatar animations with the additional help of a 3D rig optimizer model 
\cite{salvador2022}.

However, these advances in offline animation methodologies have not been extended to streaming
solutions yet. Current real-time speech-driven facial animations have employed deep neural networks
(DNNs) to prove an acoustic-visual mapping is possible \cite{realtimednn2013}, but do not offer avatar visualizations via a 3D facial mesh or latency analysis for streaming purposes.

In this work, we aim to connect the recent advances in deep articulatory inversion to improving 
real-time speech-driven facial and tongue animation. We propose a low latency streaming synthesis 
approach to predict batches of EMA data from speech using acoustic-to-articulatory inversion, animate a joint-based 
3D model in Autodesk Maya, and evaluate predicted animations by comparing the generated 
motion capture to ground truth labels. Specifically, we achieve average latencies of 130ms per 0.1 
seconds of streamed audio using shared memory buffers and demonstrate average EMA
correlations of 0.792 during real-time prediction.

\section{Methods}
\label{sec:methods}

For the proposed articulatory streaming architecture, we use an 
acoustic-to-articulatory inversion process (AAI) followed by a mapping between 
each EMA feature and a corresponding joint or curve on the 3D face model.

\subsection{Acoustic-to-Articulatory Inversion}

We first utilize articulatory inversion to generate corresponding EMA
data. This data consists of 12-dimensional features, which provide the midsagittal 
$x$ and $y$ coordinates of the tongue tip, body, and dorsum, the upper and lower lips,
and the lower incisor.

We use two models for inversion—BiGRU-based and Transformer-based. We first 
follow \cite{aai} which uses a BiGRU architecture with chunked autoregression 
and adversarial training. Specifically, we use the model with an MLP to help 
with coarticulation, a CNN as the discriminator for realistic outputs, and 
the final layer of HuBERT \cite{hsu2021hubert} to map speech into a compressed yet generalizable representation.

Additionally, we use a state-of-the-art six-layer Transformer model prepended with 
three residual convolutional blocks following \cite{gaddy2020emgspeech}. 
The model uses the tenth layer of WavLM for speech representations of audio inputs and
outputs EMA, tract variables (TVs), phonemes, and pitch simultaneously \cite{wavlm2022}.

The outputs we use from the inversion include EMA feature position ($x$, $y$) data independently
normalized to a $[-1, 1]$ range. Using the M02 speaker from the Haskins Production 
Rate Comparison (HPRC) dataset's minimum and maximum feature values, we denormalize 
the predicted EMA data into a 2D space where the covariances between features are 
preserved.

\subsection{3D Face Model}
\begin{figure}
    \includegraphics[width=\linewidth]{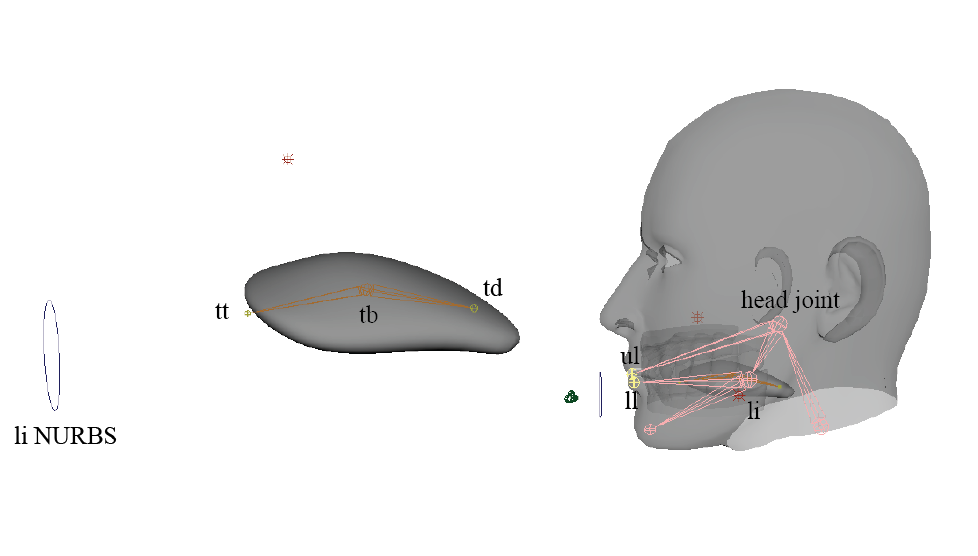}
    \caption{Midsagittal view of the 3D face avatar used for articulatory streaming; 
        tongue and NURBS curve on the left are zoomed in from full face model on the 
        right.}
    \label{fig:model}
\end{figure}

The facial model in Figure \ref{fig:model} was constructed in Autodesk Maya using teeth from \cite{teeth-model}, 
the face from \cite{face-model}, and a custom tongue. A joint-based rig was 
made for each of the tongue tip, tongue body, and tongue dorsum EMA features.

For the head, we built a similar rig with appropriate skin weights to map areas 
of the face to the corresponding joint. To approximate the hinge-based movements of
the lower incisor, we constrain the rotation of the lower incisor joint to the 
$y$-translation of a NURBS curve as controlled by the li EMA feature. The remaining
upper and lower lip features were rigged using joints connected to the head and lower
incisor joints, respectively.

When translating the $y$-dimension of the lower incisor NURBS handle, the translation 
values of the lower lip feature joint remain constant as a result of Maya's handling of 
joint rotations. To realign the EMA lower lip feature data, we calculate the global 
position of the lower lip joint when the lower incisor joint rotates:
\[
\text{ll}_x = r\cos \theta, \quad \text{ll}_y = r\sin \theta
\]
where $r$ is the radius of the lower incisor joint given by the distance between
the lower lip joint and the lower incisor joint and $\theta$ is the angle 
the lower incisor joint rotates by. Streamed EMA data then 
translates the calculated lower lip global position.

\subsection{Input Stream Processing}

During the streaming task, we use an audio input stream from a WAV file or 
from microphone input that provides waveform data in batches of 1600 samples,
which is 0.1 seconds of data for 16 kHz sampling rate audio. Since the AAI model
is not trained to infer correct EMA data for silence, we use Google's WebRTC Voice
Activity Detector (VAD) to detect whether the current batch contains speech. If not,
we show the previous frame for the next 0.1 seconds to emulate a period of silence.

If the batch contains speech, we deploy articulatory inversion to generate the
relevant EMA data. However, 0.1 seconds of audio data provides insufficient context
for accurate EMA inference, and if streamed in this manner, creates noisy animations.
To remedy this issue, we employ a rolling context window for every batch of audio
from the input stream. When we receive 1600 samples of audio data, we prepend that 
audio with the last $16000n - 3200$ samples and append the next $1600$ samples 
to the current batch to construct a window of $n$ seconds. This method creates an
intentional 100ms delay as we initially wait for two batches of audio to be 
recorded before beginning inversion to include forward context in our window. We
call the batch of data we send the ``working" batch and the data we just sent for
animation the ``sent" batch.

While this provides sufficient context to the model, we may not always have 
$n$ seconds of audio data to create a window when streaming. For example,
when streaming begins, we only have access to small amounts of audio data
being recorded. This lack of preliminary data means the model has limited
context to draw from for the first $n$ seconds of audio streamed. We try to
address this by testing four sources of initial artificial context until 
we have $n$ seconds of recorded context: silence, a recording of an articulated 
vowel, a random utterance from the HPRC dataset, and an $n$-second looped buffer 
of what data we have read so far.

The full context window is processed by the AAI model which outputs up 
to 100 frames of EMA data for the 12-dimensional features, corresponding to an 
overall 100 frames per second frequency. We discard the first 80 frames and last 10 
frames of EMA and keep only the 10 frames corresponding to the 0.1 seconds of working batch audio.

Finally, even with rolling context, the BiGRU architecture only has $n$
seconds of total audio data, and is unaware of the previous EMA outputs it
has generated. The independence from batch to batch can create inconsistencies
in the EMA data, where the end of one batch's EMA may not correspond exactly to
the beginning of the next batch's data. To smooth out these discrepancies, we 
interpolate from the last frame of the previous batch across the first four frames
of the current batch using a cubic Bézier curve. After evaluating this curve on
the first four frame steps of the current batch, we return the four interpolated 
EMA frames alongside the remaining original six EMA frames.

\subsection{Streaming to Avatar}
To stream EMA data from the AAI model to the facial model in Maya, we utilize 
concurrent processes and a shared memory buffer to facilitate fast data transfer. 
The first process initializes a shared memory buffer of 5 MB and begins processing 
the incoming audio data. Simultaneously, we use Maya's Python wrapper to begin a 
process that connects to the same memory buffer. Since Maya has low threading 
support, in this way we avoid the use of a separate thread or process to hold a 
queue for incoming data. Rather, we continuously check if the buffer has new data, 
and only continue animation if so. This also protects 
against potential packet loss issues because the shared memory buffer keeps all 
cumulative EMA data, so Maya will always have access to any data it has not animated 
yet.

After receiving a batch of EMA data in Maya, we transform every relevant joint then refresh
the Viewport 2.0, Maya's real-time hardware renderer. Thus, at every time step, all of the EMA
features will concurrently update to show their next locations, and we enable real-time 
speech-to-avatar streaming.

\begin{figure*}[!htb]
\centering
\begin{subfigure}[h]{.2\textwidth}
    \centering
    \includegraphics[width=\linewidth]{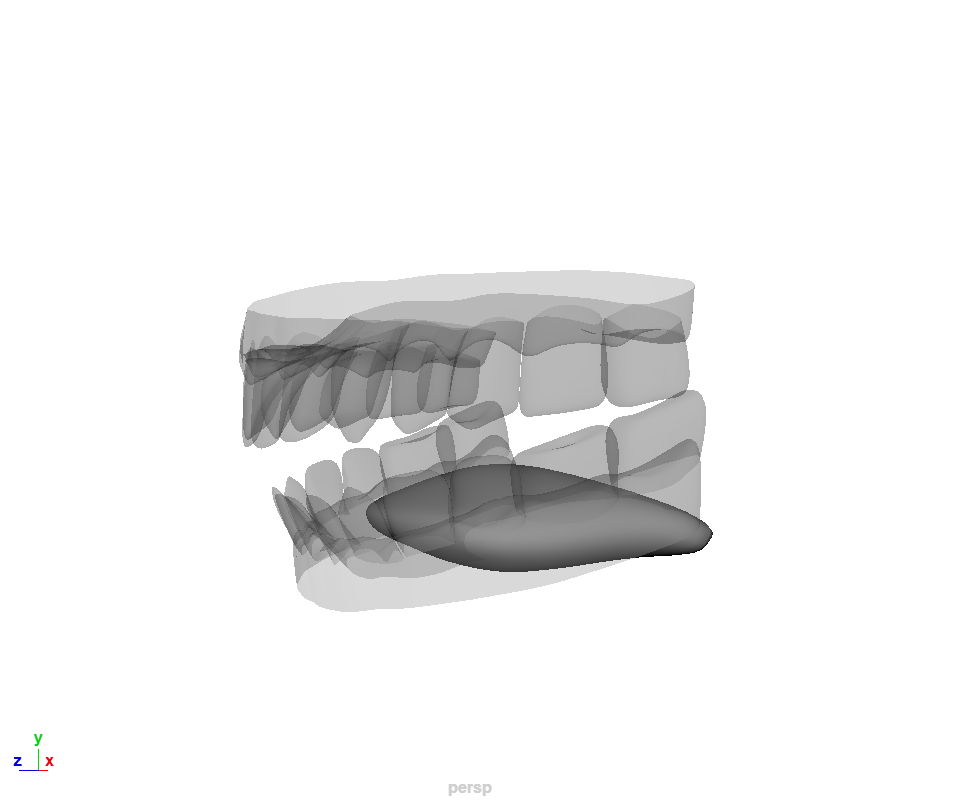}
\end{subfigure}%
\begin{subfigure}[h]{.2\textwidth}
    \centering
    \includegraphics[width=\linewidth]{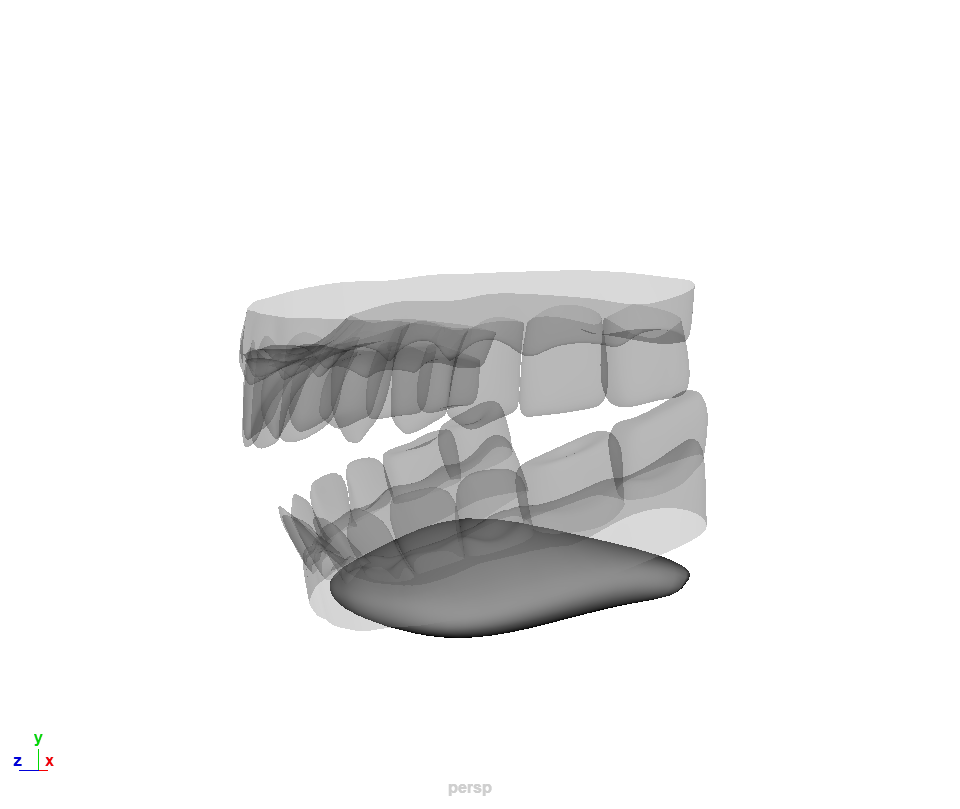}
\end{subfigure}%
\begin{subfigure}[h]{.2\textwidth}
    \centering
    \includegraphics[width=\linewidth]{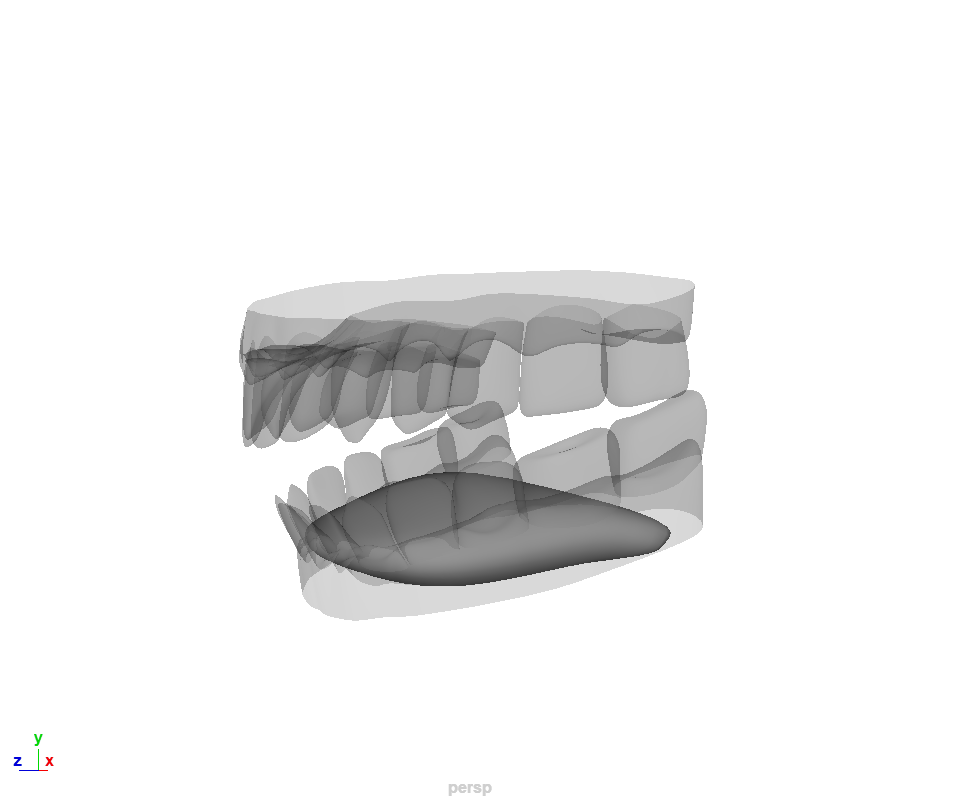}
\end{subfigure}%
\begin{subfigure}[h]{.2\textwidth}
    \centering
    \includegraphics[width=\linewidth]{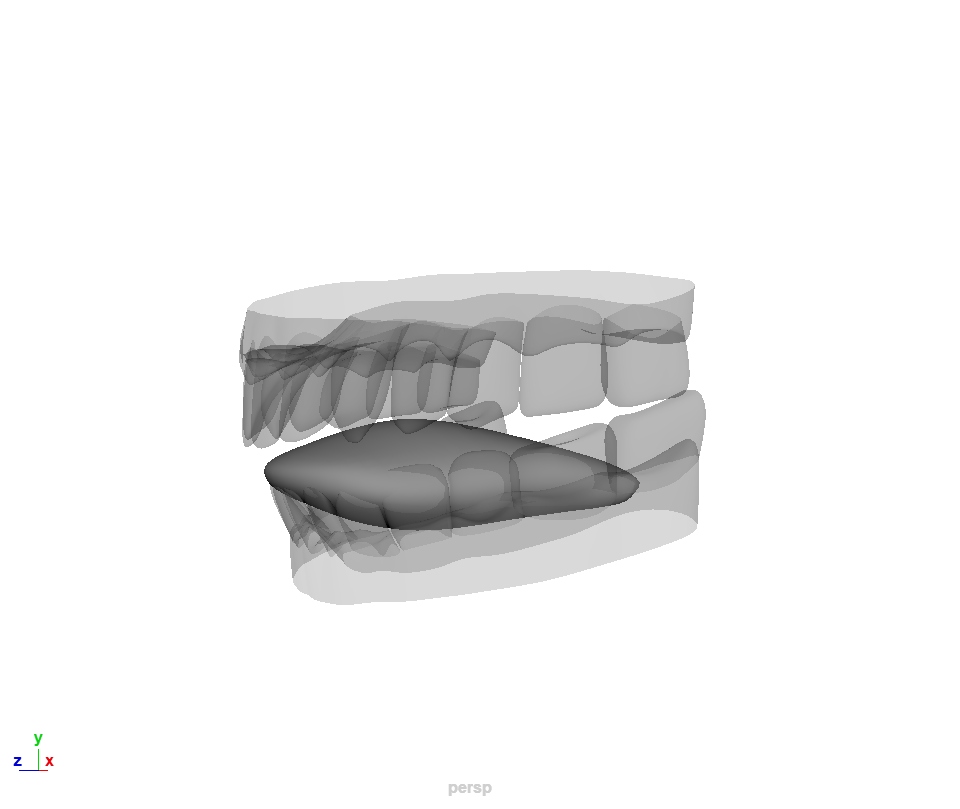}
\end{subfigure}%
\begin{subfigure}[h]{.2\textwidth}
    \centering
    \includegraphics[width=\linewidth]{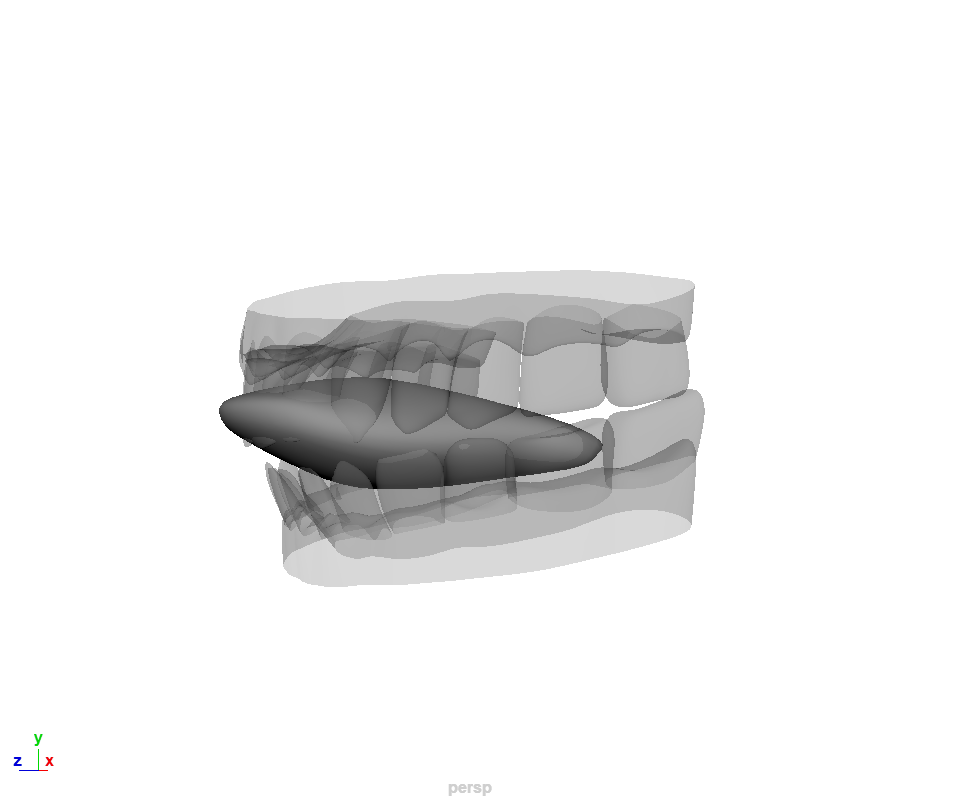}
\end{subfigure}%
\caption{``pa-ta-ka" - Series of five frames generated for audio length of 0.1 seconds}
\label{fig:visual_articulation}
\end{figure*}

\section{Results}
\label{sec:results}
\begin{figure}[ht]
    \includegraphics[width=\linewidth]{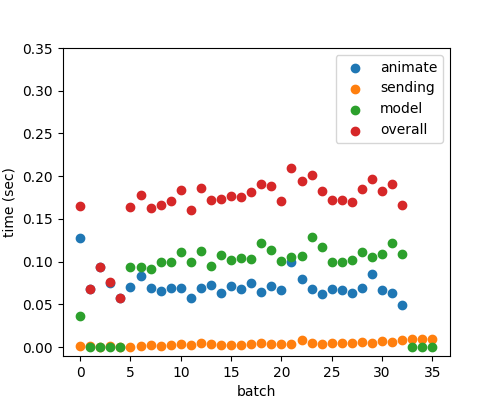}
    \caption{2-sec window profiling test for real-time performance at 100fps; each 
    batch represents 0.1 seconds of audio data, and time in seconds represents the 
    time required for a given subprocess during streaming}
    \label{fig:profiling_mouth}
\end{figure}

\begin{table}[!htb]
    \centering
    \begin{tabular}{ccc}
        \hline
        {} & \multicolumn{2}{c}{\textbf{Latency (ms)}} \\
        \textbf{Streaming Portion} & \textbf{1-sec window} & \textbf{2-sec window}\\
        \hline
        Model & 76.7 & 83.3\\
        Send & 4.19 & 4.05\\
        Animate & 56.3 & 71.8\\
        Overall & \textbf{133} & \textbf{166}\\
        \hline
    \end{tabular}
    \caption{Average latency for each portion of streaming; ``Model": articulatory inversion using BiGRU, ``Send": reading/writing the shared memory buffer, ``Animate": transforming avatar rig and refreshing the Viewport 2.0.}
    
    \label{tab:latency}
\end{table}

\begin{table}[!htb]
    \centering
    \begin{tabular}{ccc}
        \hline
        {} & \multicolumn{2}{c}{\textbf{Inversion Model PCCs}} \\
        \textbf{Artificial Context} & \textbf{BiGRU} & \textbf{Transformer}\\
        \hline
        None & 0.612 & 0.774\\
        Silence & 0.704 & 0.771\\
        Vowel & 0.705 & 0.762\\
        HPRC Utterance & 0.720 & \textbf{0.792}\\
        Looped Buffer & 0.704 & 0.769\\
        \hline
    \end{tabular}
    \caption{Average Pearson correlation coefficients for predictions
    made by each model compared to the ground truth EMA for every method
    of providing initial artificial context.}
    \label{tab:context_correlations}
\end{table}

\subsection{Streaming Latency}
Figure \ref{fig:profiling_mouth} provides an example of the latency contributions 
of each part of the streaming process. These are further summarized in 
Table \ref{tab:latency}, measuring the average latency for a 3.6 second 
length audio. When speech is detected, the largest latency bottleneck comes 
from the AAI model. The context window inversion forces the model to
convert $16000n$ samples to EMA rather than just the working 1600 samples, where
$n$ is the length of the context window. We observe the 
second largest latency contribution is from the animation of the avatar in Maya, as 
the hardware renderer requires time to refresh the viewport.

We can also see that the shared memory buffer is an effective way to stream data
when transferring data between processes on the same device, since its added latency
is very minimal even at higher size data transfers.

\subsection{Qualitative Streaming Analysis}

We visually observe real-time tongue movements accurately portraying how a real
tongue would articulate the streamed utterances. For example, Figure 
\ref{fig:visual_articulation} highlights the movement of the tongue during the ``a-ta"
portion of a ``pa-ta-ka" utterance. We qualitatively determine the tongue's position near
the front of the teeth matches how one would articulate the same utterance. 
Additionally, we observe that the avatar correctly portrays tongue positions during
prolonged vowel sounds. For example, the tongue tip nearing the lower incisor during 
an ``ah" sound or the tongue receding further towards the jaw during an ``ooh" sound
support the model's ability to make accurate predictions.

\begin{figure}[!htb]
    \includegraphics[width=\linewidth]{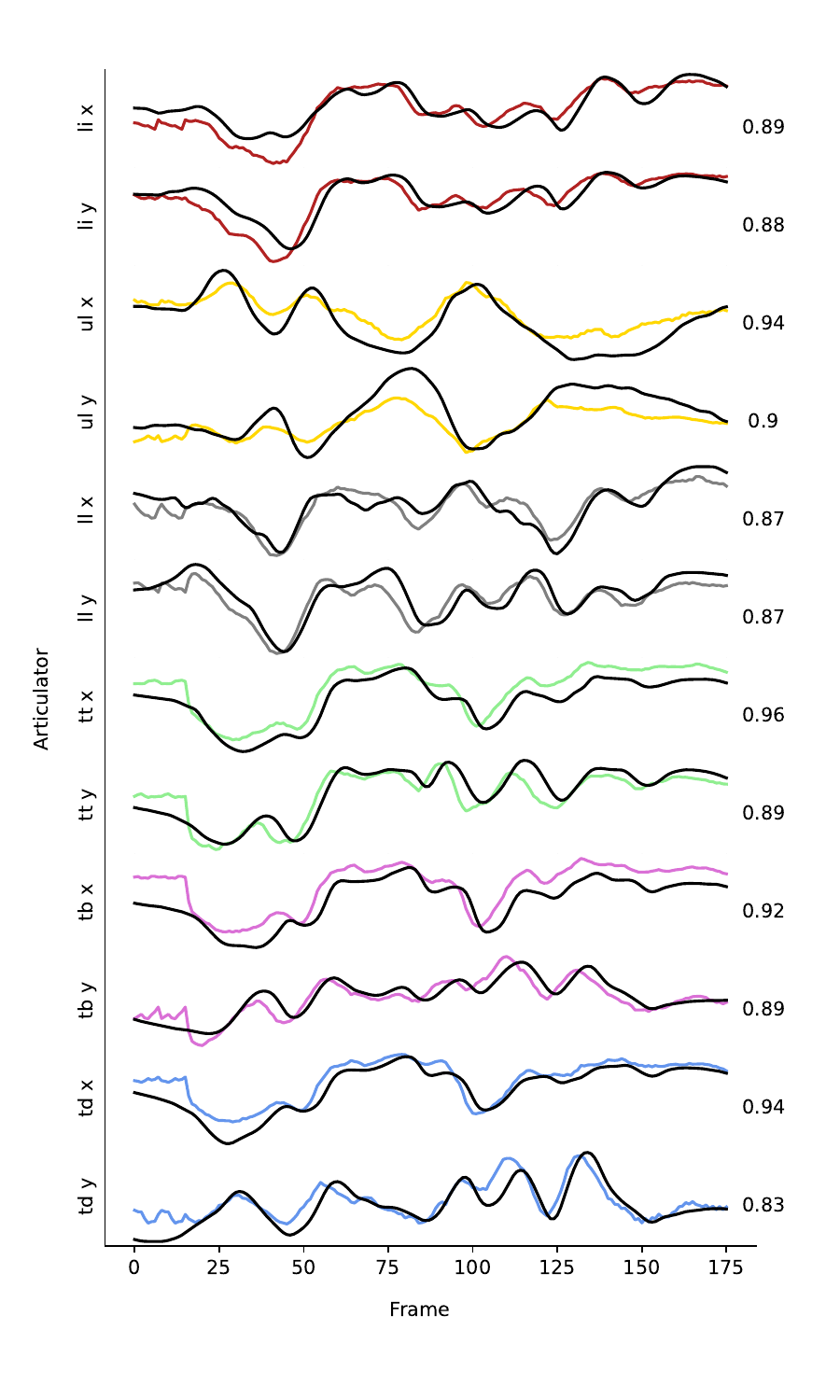}
    \caption{Midsagittal articulator movements inferred from streamed audio data using
    the Transformer-based model (in color). The trace
    of the reference EMA data also shown (in black). Pearson correlation coefficients
    (PCCs) comparing predicted trajectories to ground truth are shown to the
    right of each feature's plot.}
    \label{fig:correlations}
\end{figure}

\subsection{Streaming Articulatory Inversion Analysis}

Figure \ref{fig:correlations} highlights visual similarities between predicted and
ground truth articulator traces for each EMA feature and dimension. We observe high
Pearson correlation coefficients for each feature, markedly improving after the initial
50-100 frames. Despite best efforts to mask the lack of source data when streaming begins, 
the first 0.5-1 second of EMA data may be noisier than latter predictions. Generally in
the streaming task, the Transformer model achieves higher correlations with
the reference EMA data compared to the BiGRU model across all artificial contexts as
seen in Table \ref{tab:context_correlations}.

\section{Conclusion}
\label{sec:conclusion}

In this work, we present a real-time speech-to-avatar approach using acoustic-to-articulatory
inversion, reaching an average of 130ms latency for every batch of 0.1 seconds of audio data.
Our results enable low-latency speech-driven streaming of true-to-life mouth and tongue animation. 
We also demonstrate the efficacy of a shared memory buffer for streaming over a single device.
To the best of our knowledge, this approach is the first to facilitate real-time avatar tongue and face animations
from speech using deep learning models. We show compelling visual demonstrations of real-time
microphone-based streaming for practical use. In the future, we plan to improve the accuracy of
predicting real-time EMA data using transduction inversion techniques and eliminate the need for
manually provided context.

\section{Acknowledgements}
This research is supported by the following grants to PI Anumanchipalli --- NSF award 2106928, Google Research Scholar Award, Rose Hills Foundation and Noyce Foundation.

\vfill\pagebreak
\bibliographystyle{IEEEbib}
\bibliography{refs}

\end{document}